\documentclass[12pt]{article}
\usepackage{graphicx}
\usepackage{amssymb,amsmath,amsfonts,palatino,amsthm,epsf,epsfig}

\theoremstyle{definition}

\newcommand{\beq}{\begin{equation}}
\newcommand{\eeq}{\end{equation}}
\newcommand{\barr}{\begin{array}}
\newcommand{\earr}{\end{array}}
\newcommand{\ssz}{\scriptsize}
\begin{document}

\title{A statistical formalism of Causal Dynamical Triangulations}
\author{Mohammad H. Ansari\thanks{Email address: mansari@perimeterinstitute.ca}\  \ and
 Fotini Markopoulou\thanks{Email address:
fotini@perimeterinstitute.ca}\\
\\ \footnotesize{Perimeter Institute, Waterloo, On, Canada N2L 2Y5}
\\ \footnotesize{University of Waterloo, Waterloo, On, Canada N2L
3G1}
\\
}
\date{\today}
\maketitle
\vfill
\begin{abstract}
We rewrite the 1+1 Causal Dynamical Triangulations
model as a spin system and thus provide a
new method of solution of the model.
\end{abstract}
\vfill
\newpage

\tableofcontents

\section{Introduction}

The failure of perturbative approaches to quantum gravity has
motivated theorists to study  non-perturbative quantization of
gravity.  These  seek a consistent  quantum dynamics on the set of
all Lorentzian spacetime geometries.

One such approach which has led to very interesting results is the causal dynamical
triangulation (CDT) approach\cite{ambj98,AL}.  In the interest of understanding why
this approach leads to non-trivial results,  in this paper we study a reformulation of it as a
spin system.  The basic idea is that causal structure is coded into the values of a set of spins,
in such a way that causal relations are expressed as constraints on
the allowed spin configurations.  This makes possible a new method of studying the model which
is complementary to that used in the original papers.  In this paper we study only the
$1+1$ dimensional model\cite{ambj98},
but we believe the method described here generalizes to
higher dimensions and can include matter.

In the next section, we review the CDT model in $1+1$ dimensions and the solution
to it given by Ambjorn and Loll in \cite{ambj98}.   In section 3, we reformulate the
CDT model as a spin system, which we call the statistical model.  In section 4, we show how to
solve the model by a procedure made natural by the translation of the model into a spin system. In section 5, we discuss the application to the model of the renormalization group, after which we close with a brief discussion of what we learned about quantum Lorentzian geometry
from the translation to a spin system.

\section{Definition of the model}

In this section, we review Causal Dynamical Triangulations, as defined in \cite{ambj98}.

Causal Dynamical Triangulations is arrived at by a discretization of the path integral for quantum
general relativity in $1+1$ dimensions.
In 1+1 dimensions, the continuum Einstein action is
\beq
S[g]=\Lambda \int
\sqrt{-g}\ dA ,
\eeq
where $g=det\left( g_{ab}\right)$, and  $a, b=0,1$.   $dA$ is
the element of area  and $\Lambda$ is the cosmological
constant. The metric $g_{ab}$ represents the geometry of spacetime.

We define the amplitude to evolve from an initial geometry, which is a  circle of length
$l_0$, to a final geometry, a circle of length $l_t$, formally by
\beq
{\cal A}[l_0, l_t ] = \int  d\mu [g_{ab}] e^{iS[g]},
\eeq
where the sum is over geometries with fixed topology $S^1 \times [0,1]$,
$ d\mu [g_{ab}]$ is the measure and the boundaries of the histories are
two circles of lengths $l_0$ and $l_t$.

We now discretize this path integral.  A given history is represented by
 a set of $t$ spacelike circles (or rings), which are
labelled $S_{\left(i\right)}$. These are considered time slices in
some fixed gauge \cite{fotinilee}. Each ring has $l_{i}$ vertices,
connected by edges which are assumed to be of length unity in
Planck units. It is required that every time-slice has at least
one edge.

The vertices of two adjacent loops are connected by a set of
timelike edges of length-squared $-a^2$.  These define the causal
structure of the discrete history and are chosen so that the
surface is broken up into triangles. To ensure this,  the leftmost
future vertex of a vertex $i$-th of $S_{\left(t\right)}$ is the
rightmost future vertex of the vertex $\left(i+1\right)$-th of the
same ring. A triangle has one spacelike edge, which sits on one of
the spacelike edges of $S_{\left(t\right)}$ and two timelike
edges, which connect two vertices of $S_{\left(t\right)}$ to
either one vertex of $S_{\left(t+1\right)}$ ( ``up'' triangle), or
two vertices of $S_{\left(t-1\right)}$ (``down'' triangle). Each
history is then a piecewise flat Lorentzian geometry. In each
triangle $g=-1$, and the action becomes $S=\lambda A$, where $A$
is the summation of the areas of triangles.  The area of each
Lorentzian triangle is $\frac{\sqrt{5}}{2}a^{2}$  \cite{Sorkin
81}. Therefore, the action of a time-slice consisting of $n$
triangles is $S=\lambda n \frac{\sqrt{5}}{2}a^{2}$. We absorb the
factor of $\frac{\sqrt{5}}{2}$ in $\lambda$ and the action becomes
$S=\lambda n  a^{2}$.

The path integral amplitude for the propagation from geometry
$l_{1}$ and $l_{2}$ is the sum \beq {\cal A}[l_0, l_t ] =
\sum_{\mbox{histories}} e^{i\lambda A}, \eeq over all such
piecewise flat manifolds defined in this way with initial and
final circles fixed. Note  that  the cosmological constant
$\Lambda$ is replaced by the
 ``bare'' cosmological  constant $\lambda$.  Note also that topology change is excluded
by the requirement that each history have topology $S^1 \times [0,1]$.

In summary, the following are key assumptions of the model:

\begin{enumerate}
  \item \label{cond 1}  {\em Fixed topology}: the topology of the boundaries and
  interpolating spacelike  slices  is $S^1$ and each history is $S^1 \times [0,1]$.
  Each slice has length $\geq 1$.

  \item \label{cond 2} {\em The amplitudes are given by a path integral}: the amplitude of propagation
  from the initial ring to the  final ring is given by  the sum over all
  interpolating histories.

  \item \label{cond 3}  {\em Histories are triangulations}: the leftmost future vertex of a vertex $i$-th of $S_{\left(t\right)}$ is the rightmost
  future vertex of vertex $\left(i+1\right)$-th of $S_{\left(t\right)}$.
\end{enumerate}

\subsection{Review of the generating function method}

In this subsection we review briefly the method of solution of the
problem given in \cite{ambj98}.

Let there be $l_{t}$ vertices in $S_{\left(t\right)}$ and $l_{t+1}$ vertices in
$S_{\left(t+1\right)}$. If $k_{i}$ vertices of $S_{\left(t+1\right)}$ are in the future
of the vertex $i$ of $S_{\left(t\right)}$ then, because of condition
\ref{cond 3}, the total number of vertices of $S_{\left(t+1\right)}$ is
$l_{t+1} = \sum_{i=1}^{l_{t}}\left(k_{i}-1\right)$. The two spatial rings are
connected by $\sum_{i=1}^{l_{t}}k_{i}$ triangles; $l_{t}$ of which
are ``up''  and $l_{t+1}$ are ``down''.

To propagate from $l_0$ to $l_1$ in one time-slice, the action is
$S=\lambda a^{2} \sum_{i=1}^{l_{t}}k_{i}$  and the amplitude is

\begin{equation}
\label{eq. G1} G\left(\lambda, l_{0}, l_{1}; \Delta t=1\right) =
\frac{1}{l_{0}} \sum_{\{k_{1},...,k_{l_{0}}\}} e^{i\lambda a^{2}
\sum_{i=1}^{l_{0}}k_{i}}.
\end{equation}
Generally, if we mark the vertices of the initial ring,  the
amplitude becomes
\begin{equation}
\label{eq. G1*} G_{*}\left(\lambda, l_{0}, l_{t}; \Delta t=t\right) = l_{0}
G\left(\lambda, l_{0}, l_{t}; \Delta t=t\right),
\end{equation}
where the ${G}_*$ denotes that the vertices of the initial
ring are marked. If we mark the vertices of the
final loop, the amplitude becomes
\begin{equation}
\label{eq. G1**} G_{*}^{*}\left(\lambda, l_{0}, l_{t}; \Delta t=t\right) =
l_{0} l_{t} G\left(\lambda, l_{0}, l_{t}; \Delta t=t\right).
\end{equation}

Using the conditions \ref{cond 2} and \ref{cond
3}, the corresponding amplitude
between times $t=0$ and $t=2$ can be written as:
\begin{equation}
\label{eq. G2} G\left(\lambda, l_{0}, l_{2}; \Delta t=2\right) =
\sum_{l_{1}=1}^{\infty} G\left(\lambda, l_{0}, l_{1}, \Delta t=1\right) l_{1}
G\left(\lambda, l_{1}, l_{2},\Delta t=1\right).
\end{equation}
With marked initial vertices, the amplitude is:
\begin{equation}
\label{eq. G2*} G_{*}\left(\lambda, l_{0}, l_{2}; \Delta t=2\right) =
\sum_{l_{1}=1}^{\infty} G_{*}\left(\lambda, l_{0}, l_{1}, \Delta t=1\right)
G_{*}\left(\lambda, l_{1}, l_{2},\Delta t=1\right).
\end{equation}
Thus, we are able to write the amplitude of evolution from $t=0$ to
$t=t$ with $l_{0}$ initial vertices and $l_{t}$ final
vertices as:
\begin{equation}
\label{eq. Gt*} G_{*}\left(\lambda, l_{0}, l_{t};\Delta t=t\right) =
\sum_{l_{1}=1}^{\infty} G_{*}\left(\lambda, l_{0}, l_{1}, \Delta t=1\right)
G_{*}\left(\lambda, l_{1}, l_{t}, \Delta t=t-1\right).
\end{equation}

To solve  equation (\ref{eq. Gt*}), we Laplace transform it,
\begin{equation}
\label{laplace transform} G_{*}\left(x,y,\Delta t=t\right) \equiv
\sum_{k,l=1}^{\infty}x^{l} y^{k} G_{*}\left(\lambda,k,l;\Delta t=t\right),
\end{equation}
with the definitions $g  := e^{i \lambda a^{2}}$, $x := e^{i
\lambda_{0}a}$ and $y := e^{i \lambda_{t}a}$ in which
$\lambda_{0}$ and $\lambda_{t}$ are the cosmological constant on
the initial and final boundaries respectively. Using the Laplace
transformation (\ref{laplace transform}) on equation (\ref{eq.
Gt*}), we obtain the one-time-step $G_*$:
\begin{equation}
\label{eq. amplitude G1} G_{*}\left(x,y,g ,\Delta t=1\right) = \frac{g
^{2}xy}{\left(1 - g  x\right)\left(1 - g  x - g  y\right)}.
\end{equation}
The amplitude in terms of $g ^{m} x^{n} y^{p}$  defines the
region of convergence  $|g |\leq 1/2$, $|x|\leq 1$ and $|y|\leq 1$.
The iterative relation on
the Laplace transformed amplitude is then
\begin{equation}
\label{eq. laplace amplitude general} G_{*}\left(x,y,g ,\Delta t\right) =
\frac{g  x}{1-g  x}G_{*}\left(\frac{g }{1-g
 x},y,g ;\Delta t-1\right).
\end{equation}

Ambjorn and Loll in \cite{ambj98} showed that the iterative
relation can be written in a simpler way:
\begin{equation}
\label{eq. fixed pointed solution} G_{*}\left(x,y,g ,t\right) =
\frac{F^{2t}\left(1-F^{2}\right)^{2}xy}{\left(A_{t} - B_{t}x\right)\left(A_{t} -
B_{t}\left(x+y\right)+C_{t}xy\right)},
\end{equation}
\[ F=\frac{1-\sqrt{1-4g ^{2}}}{2g },\ \ \
A_{t}=1-F^{2t+2},\ \ B_{t}=F\left(1-F^{2t}\right),\ \
C_{t}=F^{2}\left(1-F^{2t-2}\right).
\]
The corresponding equations for the Laplace transformed amplitudes
 are:
\begin{equation}
\label{eq. G Laplace *}
G_{*}\left(x,y,g ,\Delta t=t\right) = x
\frac{d}{dx} G\left(x,y,g ,\Delta t=t\right),
\end{equation}
\begin{equation}
\label{eq. G Laplace **}
G_{*}^{*}\left(x,y,g ,\Delta t=t\right) =
y \frac{d}{dy} G_{*}\left(x,y,g ,\Delta t=t\right).
\end{equation}

The asymmetry between $x$ and $y$ in the expression (\ref{eq.
amplitude G1}) is due to the marking of the initial ring. If we
also had marked the final ring, the corresponding generating
function would be obtained from $G_{*}\left(x,y,g;\Delta t=1\right)$ by
acting with $y \frac{d}{dy}$ (which is equivalent of multiplying
$G_{*}\left(l_{0},l_{1},g,\Delta t\right)$ by $l_{1}$):
\begin{equation}
\label{eq. amp. G1**} G^{*}_{*}\left(x,y,g ;\Delta t=1\right)=\frac{g
^{2}xy}{\left(1-g  x-g  y\right)^{2}}.
\end{equation}
The corresponding generating function $G^{*}_{*}\left(x,y,g
;\Delta t=t\right)$ is obtained from \newline $G_{*}\left(x,y,g
;\Delta t=t\right)$ by acting with $y\frac{d}{dy}$,
\begin{equation}
\label {eq. Gt** fixed} G^{*}_{*}\left(x,y,g ;\Delta t=t\right)=
\frac{F^{2t}\left(1-F^{2}\right)^{2}xy}{\left(A_{t} - B_{t}\left(x+y\right)+C_{t}xy\right)^{2}}.
\end{equation}

In the continuum limit we expect that the bare propagators are
subject to a wave-function renormalization. However, all coupling
constants with positive mass dimension undergo an additive
renormalization, while the partition function undergoes a
multiplicative wave-function renormalization \cite{ambjbook}.
The only non-trivial continuum limit of eq.\ (\ref{eq. fixed pointed
solution}) is obtained when $|F|\rightarrow 1$, so $F=e^{i \alpha}$
for $\alpha \in \mathbf{R}$.  The singular Green's function can be
cured by multiplying it by a cut-off dependent factor
\cite{ambjbook}. This limit is equivalent to $|g |=\frac{1}{2
cos\alpha}$. Thus, from  the convergence condition, we find
$g =\pm 1/2$ (at $\alpha=0 , \pi)$.

\section{The dual statistical model}

We now recast the 1+1-dimensional Causal Dynamical Triangulations
model as a spin system with certain constraints on their
configurations and couplings, reflecting the geometric properties
of the CDT.\footnote{Another statistical mechanical approach to
the cdt model is developed in \cite{stat}" }

We proceed as follows.  Each triangle $i$ will be regarded as a
spin $\sigma_i$. We associate to each down triangle (with two
vertices on $S_{\left(t\right)}$ and one on
$S_{\left(t-1\right)}$) an up spin and to each up triangle  (with
two vertex on $S_{\left(t\right)}$ and one on
$S_{\left(t+1\right)}$) a down spin. Spins can take two possible
values, $\sigma_+=gx$ when they come from an up triangle, and
$\sigma_-=gy$ when they correspond to a down triangle.  The spins
will live on a trivalent lattice and we find it convenient to
graphically denote the two states of a spin as: \beq \label{eq.
spin up and down} \sigma_i:=\left\{ \barr{lll} \sigma_+ = gx,
&\barr{c}
\includegraphics[height=2cm]{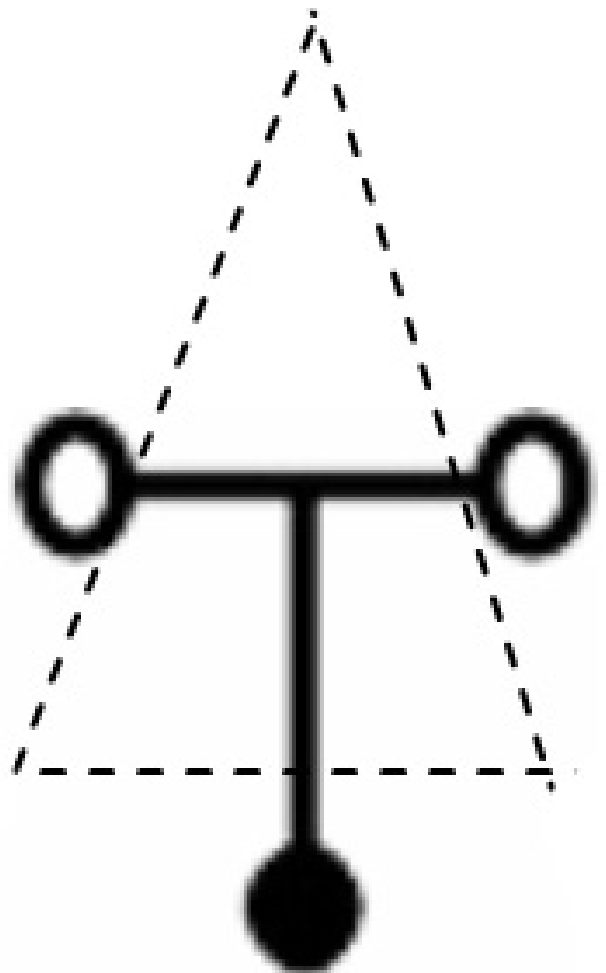}
\earr&{\mbox{``spin down'',}}\\
&&\\
\sigma_- = gy, &\barr{c}
\includegraphics[height=2cm]{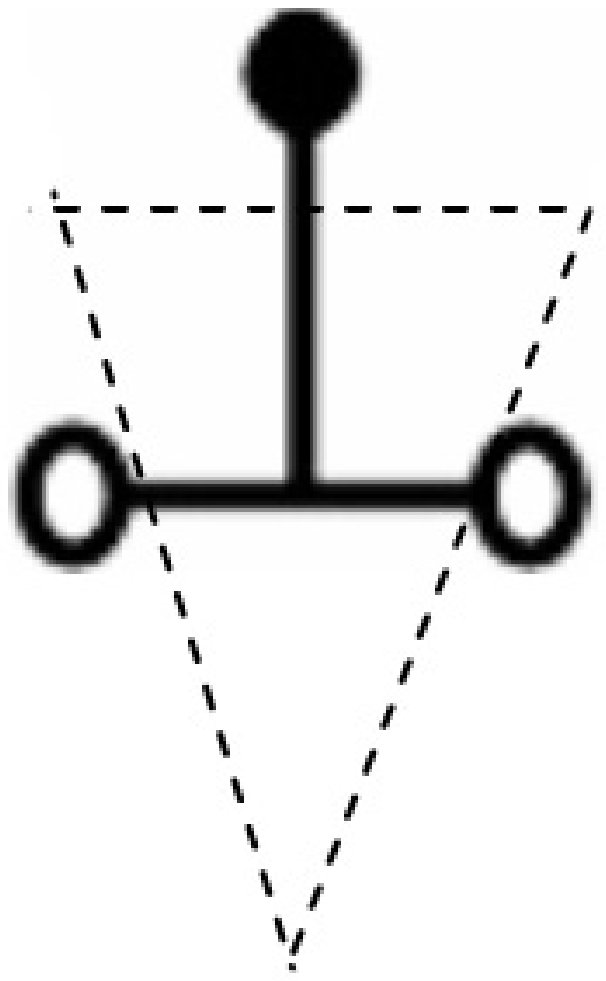}
\earr&{\mbox{``spin up''.}} \earr \right. \eeq The dashed line
represents the triangle dual to each spin. Each spin has one black
head and two white head, which are dual to one spacelike edge and
two timelike edges, respectively.

Gluing two triangles along a common edge defines a spin-spin
interaction. The CDT weighing of the triangulations means there
are two kinds of interactions: gluing two triangles along their
spacelike edges gives a spin-spin interaction of strength $J_S$,
while a gluing of two timelike edges corresponds to coupling $J_T$
as follows:

\beq \label{eq. Js and JS} J:=\left\{ \barr{rll}
J_S&=\frac{1}{xy},&\qquad{\mbox{spacelike}}\\
J_T&=1,&\qquad{\mbox{timelike.}} \earr \right.  \eeq

No gluing of a timelike to a spacelike edge is allowed. We have
incorporated these couplings to the graphical notation (\ref{eq.
spin up and down}): the interaction between two black heads of two
spins occurs with the spin-spin interaction coupling $J_S$, and
the interaction between two white heads does with a $J_T$
coupling.\footnote{Sometimes in this paper we call $J_T$ coupling
the ``white interaction'' because of the coupling it provides
between two white heads, and $J_S$ coupling the ``black
interaction'' because of the coupling it provides between two
black heads.}

In an ordinary spin system, the values of spins
are not related to the structure of the lattice. The
lattice is fixed while the values of the spins on the nodes vary.
However, in a model of gravity we do not have any pre-assigned
lattice, since the spacetime is completely dynamical. In our
formalism spacetime is created by the configurations of the spins.

\begin{figure}[h]
\begin{center}
\includegraphics[height=2.5cm]{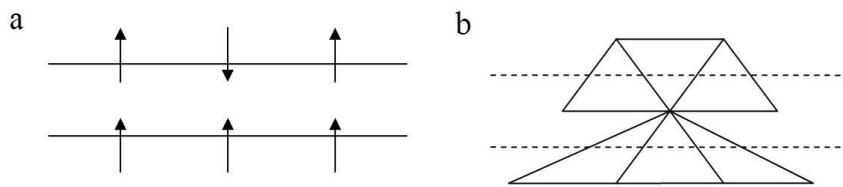}
\caption{a) An arbitrary spin configuration and b) the dual
geometry}
\end{center}
\end{figure}

Let us look at a simple example to explain how the spins relate to
the causal structure and geometry.  In Figure 1(a), we see a
lattice consisting of  two rows with a certain number of spins. In
an ordinary spin system without an external field, the spins
fluctuate independently of the lattice. However, in our model, the
spins define  the causal structure of the resulting geometry.
Following the rules just described,  Figure 1(a) is interpreted as
a dual geometry, shown in Fig. 1(b). We see from Figure 1(b) that
the spins in Fig. 1(a) define a geometry that does not satisfy the
causality constraints of CDT. The dual spacetime is  not causal as
there are  vertices in the second slice that have no past in the
initial slice.

We will impose constraints on the dual spin system so that all
spin configurations have dual CDT histories:
\begin{enumerate}
\item{\em Causality constraint:} Each up spin in row $t$ ($t<t_{final}$) must be coupled
to a down spin in row $t+1$, with coupling   $J_S$.
\item{\em Non-degeneracy constraint:}  Every row has at least one spin.
\end{enumerate}

A spin system that satisfies these constraints is dual to a
history of the CDT model. \footnote{From the point of view of
statistical physics, the present model is analogous to
a``diluted'' 2 dimensional classical Ising model because the
number of spins in each time-slice can vary (see, for example,
\cite{diluted}).}

It is instructive to give an example of a spin configuration that
does satisfy the causality constraints.  The triangulation \beq
\includegraphics[height=2cm]{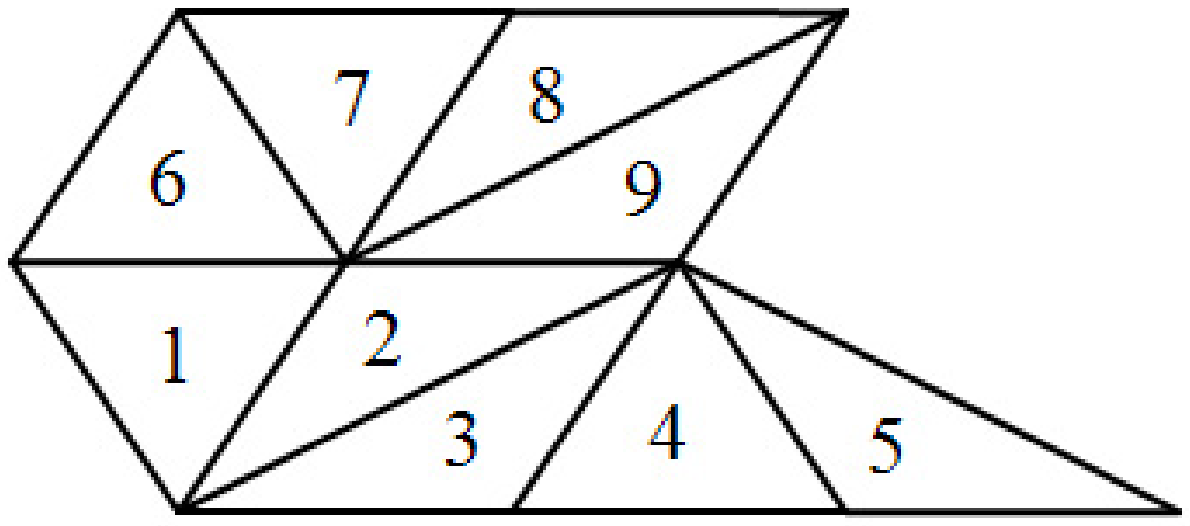} \eeq
 in the dual  spin system is:
\beq \label{graph example}
\includegraphics[height=4cm]{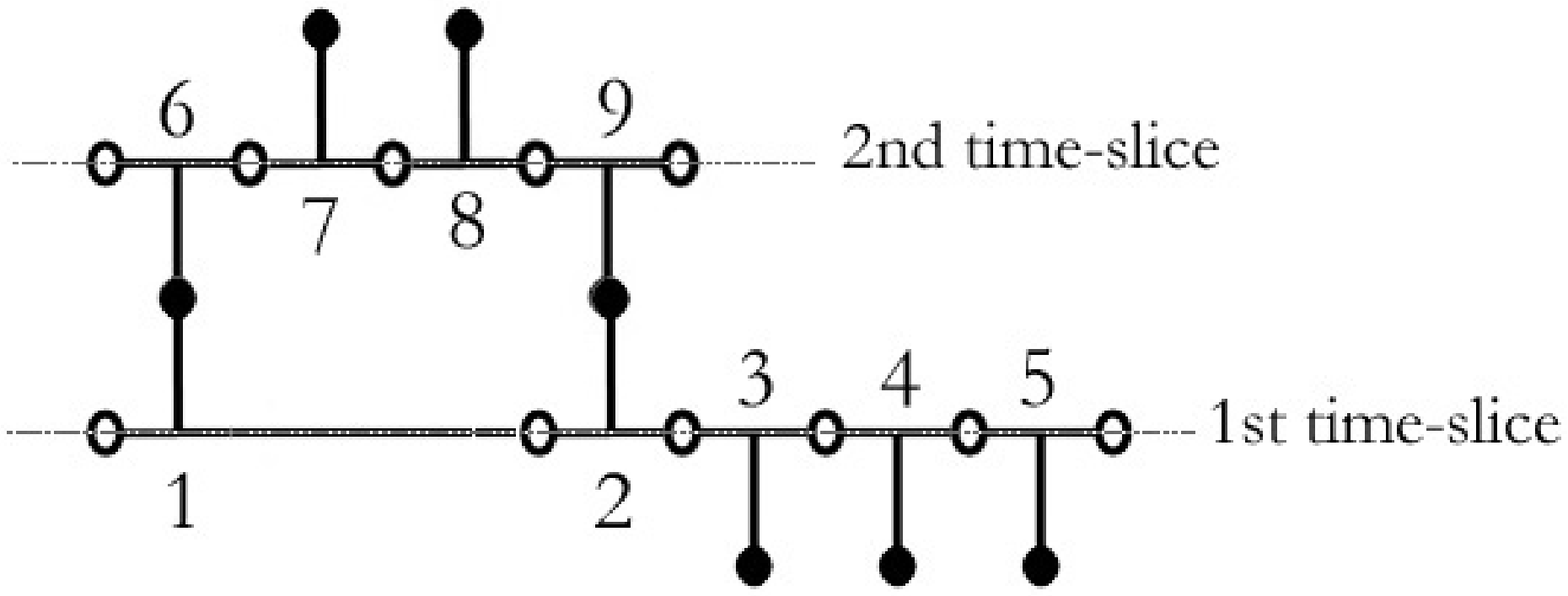} \eeq
The amplitude of this history  is:
\beq
A= \left(\sigma_{1}
J_{S} \sigma_{6}\right) \ J_{T} \ \sigma_{7} \ J_{T} \ \sigma_{8} \
J_{T} \ \left(\sigma_{9} J_{S} \sigma_{2}\right) \ J_{T} \ \sigma_{3} \ J_{T}
\ \sigma_{4} \ J_{T} \ \sigma_{5} = g ^{9} x^{4} y^{2}.
 \eeq
Note that the dual spin system can be read as a history somewhat
analogous to a Feynman diagram.  The history of (\ref{graph
example}) begins with three initial black circles, and ends with
two final black circles.  In between there are two moments of
discrete time, given by the rows of $J_T$ couplings.  There are no
unpaired black or white circles, except for the initial and final
black ones (recall that we have cylindrical boundary conditions on
the boundary white ones).\footnote{An example of the spacetime
structure with equal intermediate rings can be thought of as the
tower of Pizza (from above of which Galilei did his famous
gravitational experiment).}

Next, we use this definition of the model to count the 2d CDT
histories.

\section{Computing CDT amplitudes using the dual spin system}

We now want to calculate the 1+1 CDT path integral amplitude from a given initial ring
to a given final ring, using the dual spin system.  Here,  the evolution amplitude is given by a
correlation function between the spins of the boundary rows.

We will find it convenient to introduce a notion of effective
spins, which allows us to sum the causal histories.

\subsection{Single-row configurations}

We will illustrate the spin system solution by first calculating the CDT amplitude for a single row of spins.  This means
all spin configurations on rows containing from one to infinite number of spins, subject
to the causality and non-degeneracy constraints.

In fact, for one row, we only have the non-degeneracy constraint, which we
 impose  using what
we call ``the box''. A box is two spins with opposite orientations, with one $J_T$ coupling.
This is denoted by
 \beq \label{graph A2}
 \barr{c}\mbox{\epsfig{figure=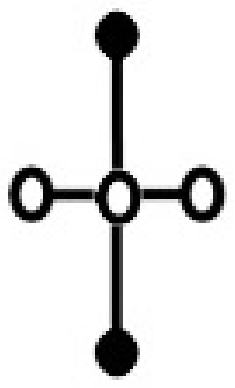,height=1.5cm}}\earr:= A^{\left(1\right)}_{\mbox{\ssz box}} =
 \frac{1}{2}\left(\barr{c}\mbox{\epsfig{figure=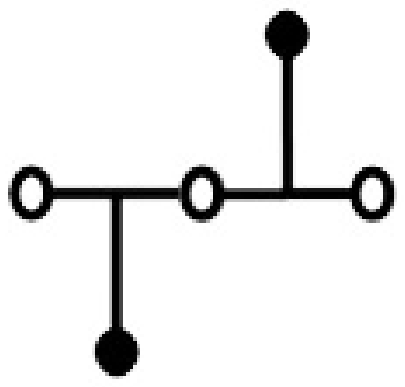,height=1.5cm}}\earr +
 \barr{c}\mbox{\epsfig{figure=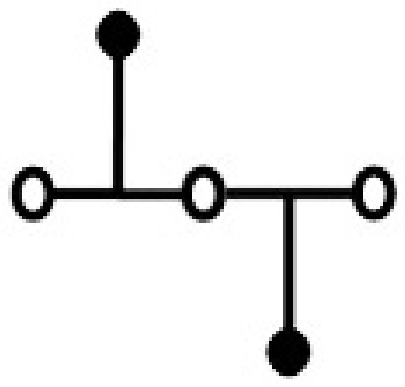,height=1.5cm}}\earr\right)
= g^2xy. \eeq
The superindex
$^{\left(1\right)}$ is a reminder that we are working with one-row
configurations. The geometric constraint is equivalent to
requiring that each row must contain {\em at least one box}.

For a single row of spins, there are no $J_S$ couplings and we
have simply the combinatorics of spins attached to a box. In a row
of $N$ spins,  two of them have to make a box, and are no longer
free to be up or down\footnote{Since the spins are located on a
closed chain, the amplitude is not sensitive to locations of
spins. On the other hand since the free black heads on the initial
and final rings are marked, different location of $A_{box}$ among
spins is counted as a different configurations.}, hence, \beq
A^{\left(1\right)}_{N}=c_{N}A^{\left(1\right)}_{box}
\left(\sum_{i=1,2} \sigma_{i}\right)^{N-2} , \eeq where $c_{N}$ is
the number of ways of choosing two consecutive marked spins (the
box) among $N$ spins, $c_{N}=\frac{2}{N} \left(^{\ \
N}_{N-2}\right)=N-1$.

For example, for $N=4$:

\[A^{\left(1\right)}_{N=4}=
 \barr{c}\mbox{\epsfig{figure=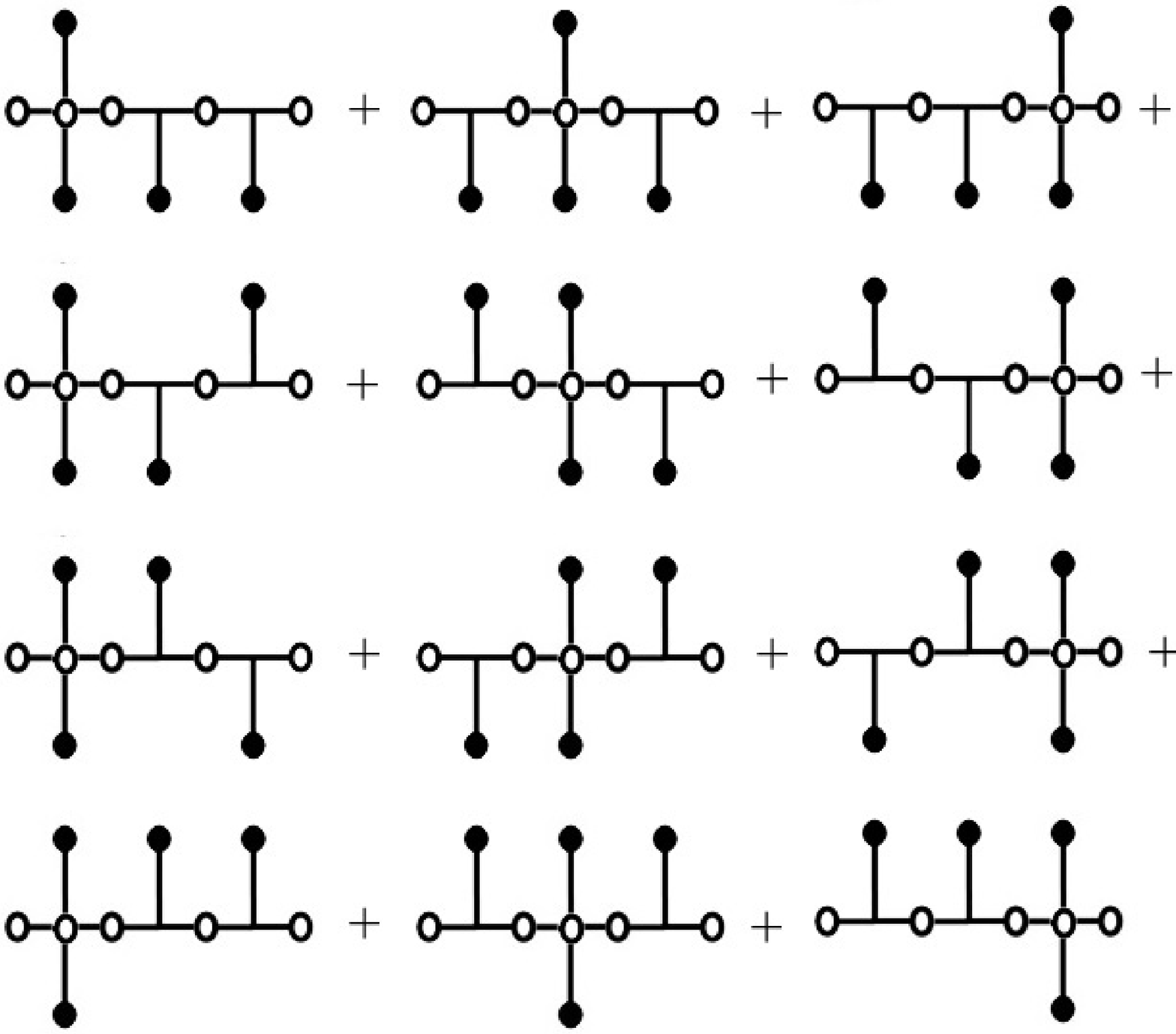,height=7cm}}\earr\]
\beq = 3 A^{\left(1\right)}_{\mbox{\ssz box}} \left(\sum_{i=1,2}\sigma_{i}\right)^{2} =
3\left(g^2xy\right)\left(gx+gy\right)^{2}. \eeq

This amplitude  is actually the Laplace transformed
amplitude, similar to a Greens function.
In the dual triangulation, the Laplace inverse transform of the above is the amplitude to go from an initial ring of length $l_{0}$
to a final ring of length $l_{1}$.  So we
have to add up all Laplace transformed amplitudes for different ring lengths  (spin number and configurations) to derive the
Green's function.

The single-slice amplitude, $A^{\left(1\right)}$, is \beq
A^{\left(1\right)}=\sum_{N=2}^{\infty}A^{\left(1\right)}_{N}=A^{\left(1\right)}_{\mbox{\ssz
box}}+g^2xy \sum_{N=3}^{\infty}\left(N-1\right)g^{N-2}\left(x+y\right)^{N-2}. \eeq
To perform the sum, we redefine $n=N-2$ to obtain:
\begin{eqnarray}
A^{\left(1\right)}&=&g^2xy\left[1+\sum_{n=1}^\infty  \left(n+1\right)g^n\left(x+y \right)^n\right]\nonumber\\
    &=&g^2xy\sum_{n=0}^\infty\left(n+1\right)g^n\left(x+y\right)^n.
\end{eqnarray}
 We now define the quantity $A:=g\left(x+y\right)$, so that the above
equation can be rewritten as \beq
A^{\left(1\right)}=g^2xy\sum_{n=0}^\infty\left(n+1\right)A^n.
\label{eq:A1} \eeq For $|A|\leq 1$, eq.\ (\ref{eq:A1}) sums to the
same result of (\ref{eq. amp. G1**}): \beq
A^{\left(1\right)}=\frac{g^2xy}{\left(1-A\right)^2}=\frac{g^2xy}{\left(1-gx-gy\right)^2}.
\label{eq:A26} \eeq

\subsubsection{Non-marked vertex amplitude}

Comparing the amplitude (\ref{eq:A26}) with the results of
\cite{ambj98}, we see our result is symmetric on $x$ and $y$ since
both our up and down spins are marked. To derive the exact result
of Ambjorn and Loll in \cite{ambj98} we need unmarked up spins.

We can illustrate this for  $N=4$:
\[A^{(1)}_{*N=4}=\barr{c}\mbox{\epsfig{figure=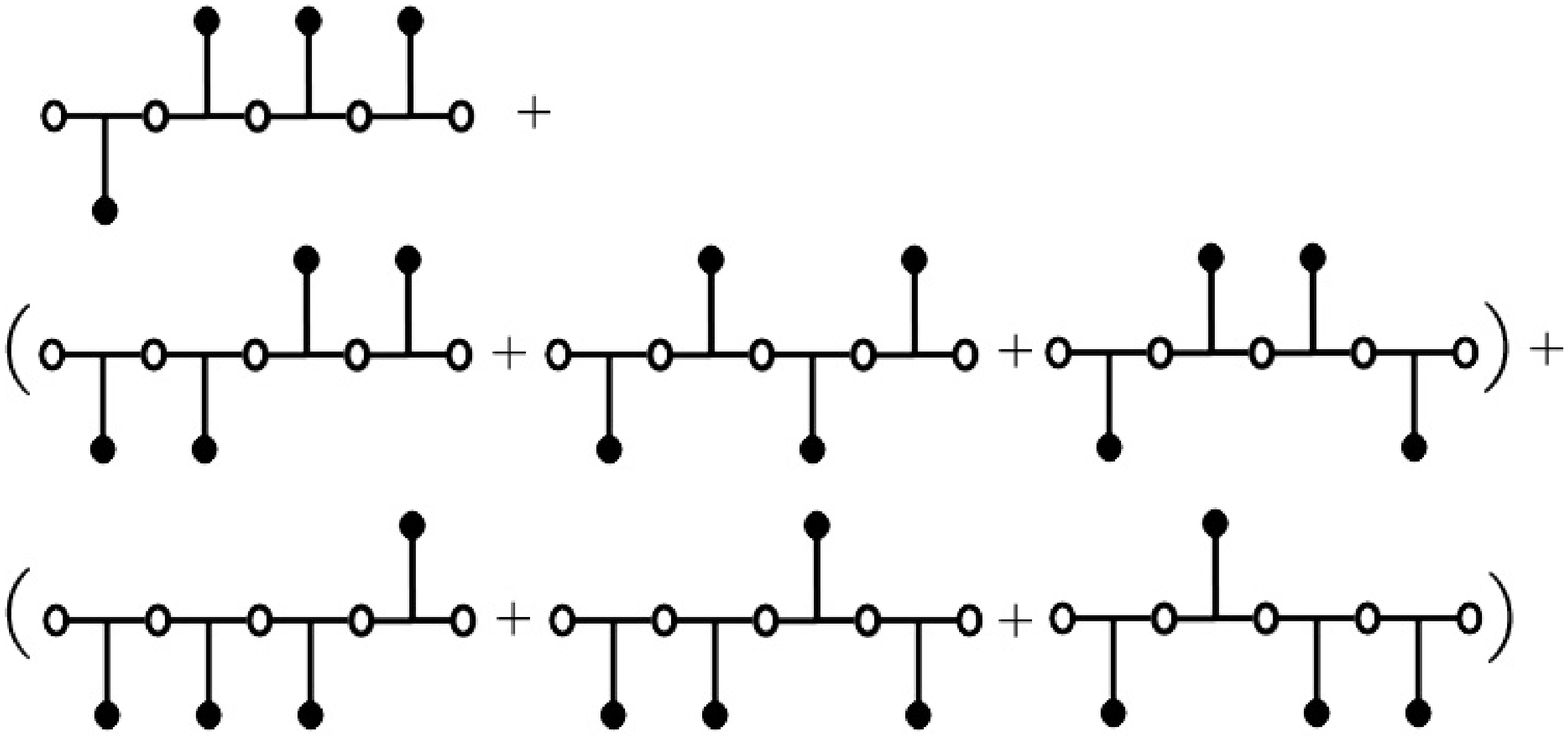,height=4.5cm}}\earr\]
\beq = g^{4}xy\left(y^2+3xy+ 3x^2\right).\eeq
Since
the top black circles are not distinguished, we have only one term with 3 up spins.
For two up spins there are three terms, for the possible permutations of the two marked down spins, etc.

Summing  over all spins, we find the unmarked amplitude
$A_*^{(1)}$ for one row to be:
\begin{eqnarray}
A_*^{(1)}&=&
\barr{c}\mbox{\epsfig{figure=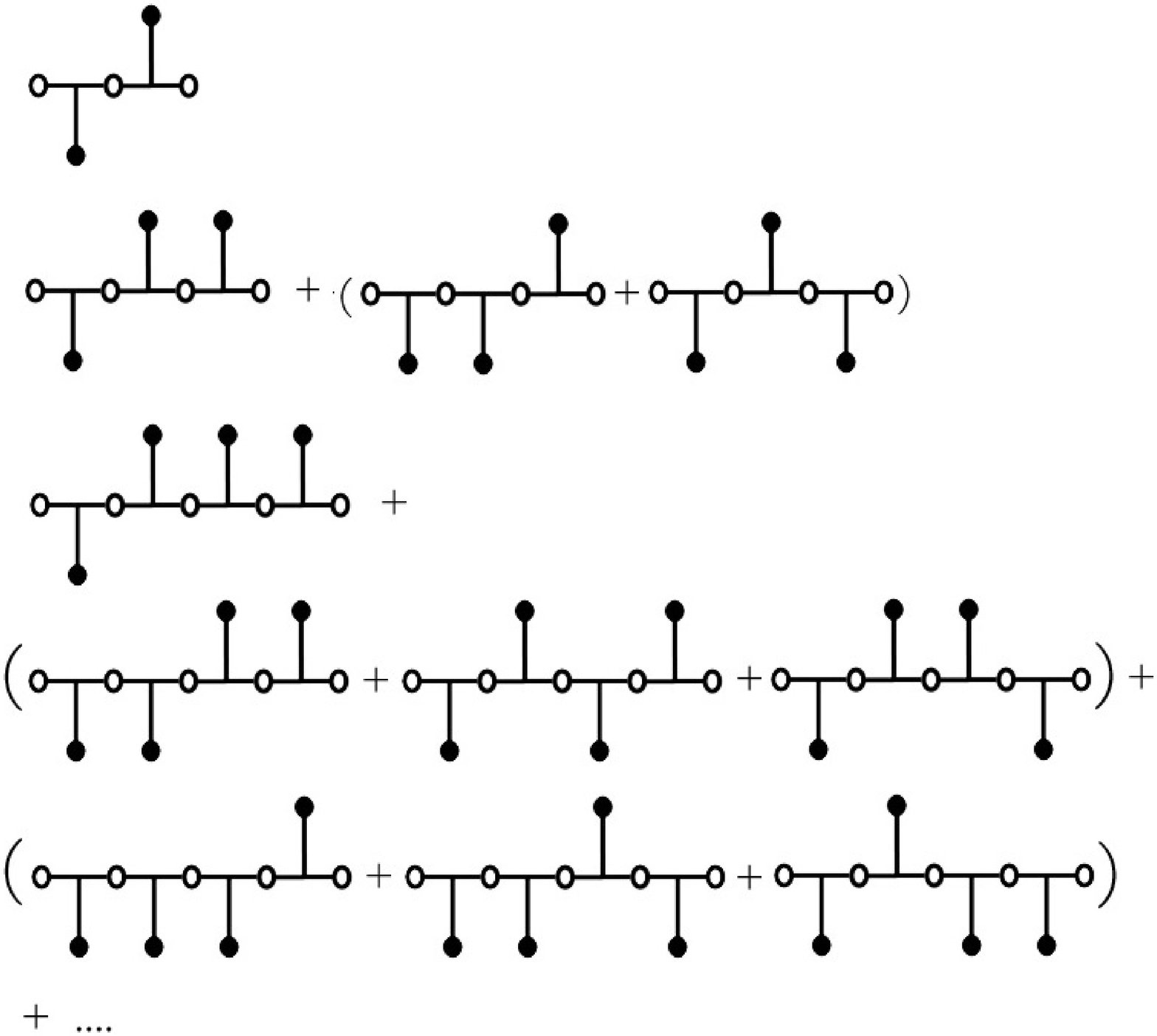,height=8.5cm}}\earr\nonumber\\
&=&
g^2xy+g^2xy\left(gy+2gx\right)+
g^4xy\left(y^2+3xy+ 3x^2\right)+\ldots .
 \label{eq. A_*^1 (first)}
\end{eqnarray}
To sum the infinite series we use a simple trick. We add and
subtract $1+gx+g^2x^2+g^3x^3+\ldots$ from the amplitude (\ref{eq.
A_*^1 (first)}) to find
\begin{eqnarray}
A_*^{(1)} &=&\left(1 + gx\left( 1 + gy + g^2y^2 + \ldots\right) +
g^2x^2 \left( 1 + gy + g^2y^2 +
\ldots\right)^2 + \ldots \right)\nonumber \\
&&-\left( 1 + gx + g^2x^2 + g^3x^3 + \ldots\right)\nonumber\\
&=& \frac{g^2xy}{\left(1-gx\right)\left(1-gx-gy\right)},
\end{eqnarray}
which is the same result of the unmarked one-step transfer matrix
amplitude (\ref{eq. amplitude G1}). It is easy to see that using
the eq. (\ref{eq. G Laplace **}),

\begin{equation}
\label{eq. A* and A} A^{(1)} = y \frac{d}{dy} A_{*}^{(1)}.
\end{equation}

In fact $A^{(t)}(x, y, g)$ (the corresponding amplitude of marked
initial and final spins in $t$ slices) is exactly equivalent to
the $G_*^*(x, y, g, t)$ solution of the generating function
method.

\subsection{Two-row configurations:  the effective spin
$\Sigma^{\left(2\right)}$}

Our solution of the CDT departs from the transfer matrix method of
\cite{ambj98} when more than one row is considered.  For a spin
configuration of $n$ rows of spins, we will introduce an effective
spin $\Sigma^{\left(n\right)}$ which makes the $n$ rows of spins
$\sigma_i$ a single row of these effective spins.  Once the
appropriate form of the effective spin has been found the
remaining calculation is very straightforward. \footnote{Let us
remind that any horizontal sequence of black heads make a
time-slice as well as any horizontal sequence of white heads.}

The form of the effective spin $\Sigma^{\left(n\right)}$ is different for odd or even $n$.
We start with the 2-row effective spin $\Sigma^{\left(2\right)}$.   This effective spin has the form
\beq
\label{eq. spin2 diagram}
\Sigma^{\left(2\right)}=
\barr{c}\mbox{\epsfig{figure=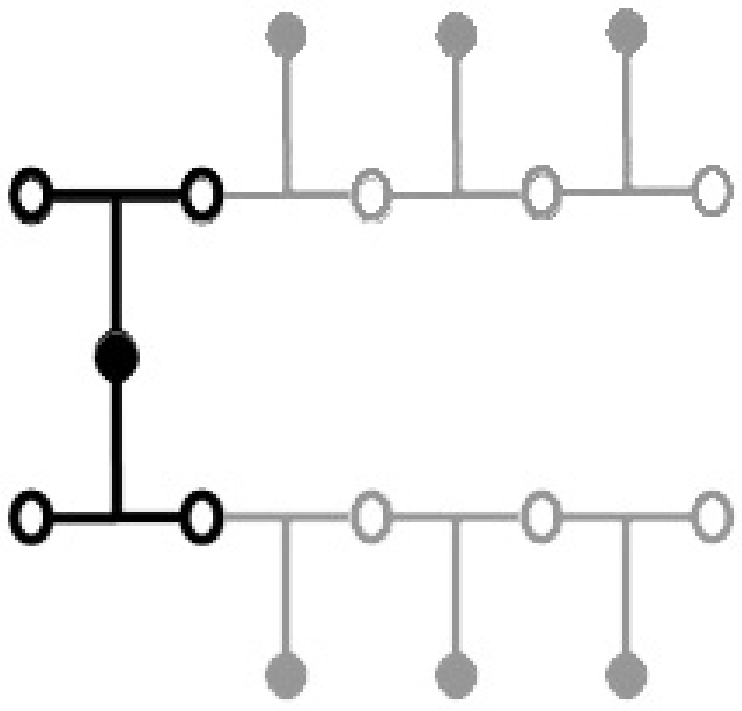,height=3cm}}
\earr
\eeq
where the gray spins denote rows with spins from zero to infinity so that the above diagram is
 \begin{eqnarray}
\Sigma^{\left(2\right)} &= &\left(gx\right)\ J_S \left(gy\right)
\sum_{l_{0}=0}^{\infty}\left(gx\right)^{l_{0}}\sum_{l_{1}=0}^{\infty}\left(gy\right)^{l_{1}}\\
&=&\frac{g^2}{\left(1-gx\right)\left(1-gy\right)}=\frac{g^2}{\left(1-\sigma_+\right)\left(1-\sigma_-\right)}.
\label{eq. effective spin 2}
\end{eqnarray}
Note that the solid black part of $\Sigma^{\left(2\right)}$ is the implementation of the causality constraint for two rows.

To calculate the two-row amplitude $A^{\left(2\right)}$, we first
need $A^{\left(2\right)}_{box}$. The diagram of
$A^{\left(2\right)}_{box}$ differs from $\Sigma^{\left(2\right)}$
in that in $\Sigma^{\left(2\right)}$ the number of the gray spins
run from zero to infinity while in $A^{\left(2\right)}_{box}$ they
run from one to infinity. In other words, the presence of at least
one up spin  in the final row and one down spin in the initial row
is guaranteed.  However, the end row spins are distinguishable and
have to be counted separately. We thus  easily find the amplitude
for $A^{\left(2\right)}_{box}$ to be: \beq \label{eq. A^2_box}
A^{\left(2\right)}_{box}= \left(gx\right)J_S\left(gy\right)
\sum_{l_{0}=1}^{\infty}l_0\left(gx\right)^{l_{0}}\sum_{l_{1}=1}^{\infty}l_1\left(gy\right)^{l_{1}}
=
\frac{g^4xy}{\left(1-\sigma_-\right)^2\left(1-\sigma_+\right)^2}.
\eeq

The amplitude  $A^{\left(2\right)}$ is obtained from all possible configurations of effective spins attached to one
$A^{\left(2\right)}_{box}$. Since the spins are marked, as in the amplitude of
the one-row case, the position of $A^{\left(2\right)}_{box}$ has to be taken into account.   The final result is:
\beq \label{eq. A2} A^{\left(2\right)}=A^{\left(2\right)}_{box} \sum_{N=0}^{\infty}
\left(N+1\right)\left(\Sigma^{\left(2\right)}\right)^N = \frac{ A^{\left(2\right)}_{box}}{\left(1-\Sigma^{\left(2\right)}\right)^2}.
\eeq

\subsection{The odd and even effective spins}

We can now generalize the method we used to count one- and two-row
Green's functions to the general case. We shall find it useful to
consider separately the cases of even and odd numbers of slices.
Since there is no interaction between a white and a black head, we
can divide interactions and cover all black interactions by
defining the notion of effective spin, and white-interact them
along their common time-slice (-slices).

Note that, in the
one-row case, there were two values of spin
$\sigma$, up and down. In the two-row case, there is only
one type of effective spin. This generalizes: even-row configurations
can be mapped to single-row with one type of effective spin, while
odd-row configurations need two-valued effective spins.

\subsubsection{Odd effective spins}

It is instructive to find the effective spin for three rows first.
One can see that all three-row configurations can be constructed from these two building blocks:

\beq \label{eq. spin3 up and down} \Sigma^{\left(3\right)}:=\left\{
\barr{lll} \Sigma^{\left(3\right)}_+ = \frac{g^2}{1-\sigma_+}, &\barr{c}
\includegraphics[height=2cm]{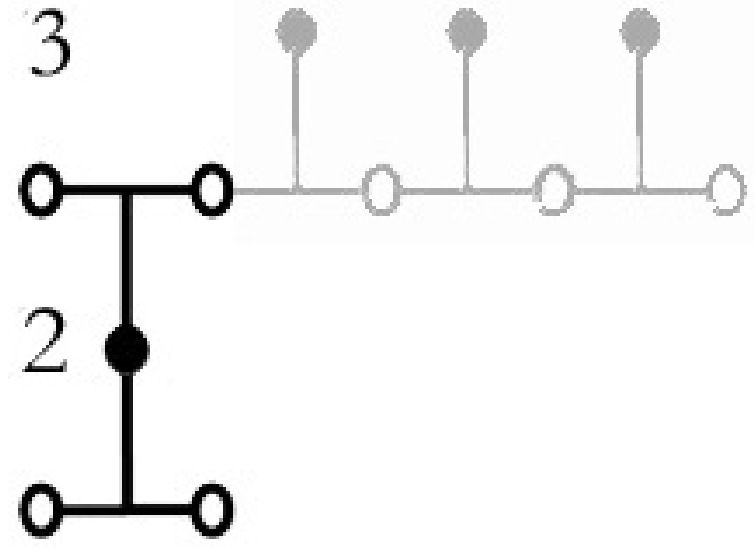}
\earr&{\mbox{``three-row effective up spin''}}\\
&&\\
\Sigma^{\left(3\right)}_- = \frac{g^2}{1-\sigma_-}, &\barr{c}
\includegraphics[height=2cm]{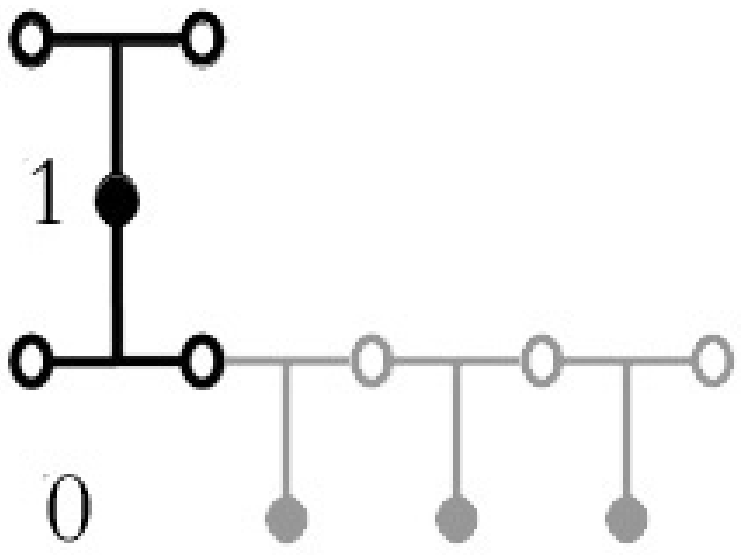}
\earr&{\mbox{``three-row effective down spin''.}} \earr \right.
\eeq

The labels 0, 1, 2 and 3 on the black heads denote the order of
the rows, from initial, 0, to final, 3.  There are two different
sequences of $\sigma_i$ spins in
 each one of the
effective spins.  $\Sigma^{\left(3\right)}_+$ has a sequence of
$\sigma_+$ spins in row 3, while $\Sigma^{\left(3\right)}_-$ has a
sequence of $\sigma_-$ spins in row 1. Again, the number of gray
spins varies from one to infinity.

$\Sigma^{(3)}_-$ and $\Sigma^{(3)}_+$ can white-interact by their
common white heads. In fact, any the connection between the
initial slice (the zeroth row) and the final slice (the 3rd row)
occurs if and only if at lease one $\Sigma^{(3)}_-$
white-interacts with at least one $\Sigma^{(3)}_+$. This minimal
interaction makes $A^{(3)}_{box}$.

We evaluate $A^{\left(3\right)}_{box}$ as previously. We need at
least one $\Sigma^{\left(3\right)}_+$ and one
$\Sigma^{\left(3\right)}_-$ in each configuration. The initial and
final loops all spins are marked and thus it matters which one of
the  gray spins we convert to a black spin (whose presence is
necessary in $A_{box}$). Thus, like (\ref{graph A2}) we have:

\begin{eqnarray}
A^{\left(3\right)}_{box} &=& \frac{1}{2} \left(g^2\
\sum_{l_0=1}^{\infty} l_{0} \left(gx\right)^{l_{0}} \right) J_T
\left(g^2\ \sum_{l_3=1}^{\infty} l_{3}
\left(gy\right)^{l_{3}}\right) \nonumber\\
&+& \frac{1}{2} \left(g^2\ \sum_{l_3=1}^{\infty} l_{3}
\left(gy\right)^{l_{3}}\right) J_T
\left(g^2\ \sum_{l_0=1}^{\infty} l_{0} \left(gx\right)^{l_{0}}\right) \nonumber\\
&=& \frac{g^6xy}{\left(1-\sigma_-\right)^2\left(1-\sigma_+\right)^2}\nonumber\\
&=& g^2 A^{\left(2\right)}_{box}. \label{eq. A3box}
 \end{eqnarray}
The three-row amplitude  is now easily
derived from the possible configurations of $N$ effective spins  on a
single  $A^{\left(3\right)}_{box}$:
\beq \label{eq. A3} A^{\left(3\right)} = A^{\left(3\right)}_{box} \sum_{N=0}^{\infty}
\left(N+1\right) \left(\Sigma^{\left(3\right)}_++\Sigma^{\left(3\right)}_-\right)^{N} =
\frac{A^{\left(3\right)}_{box}}{\left(1-\Sigma^{\left(3\right)}_+-\Sigma^{\left(3\right)}_-\right)^2}.
\eeq

Generalizing this, it is straightforward to check that, for odd-row configurations,
$\Sigma^{\left(j\right)}_\pm$,
 $A_{box}^{\left(j\right)}$ and amplitude $A^{\left(j\right)}$ are:
\beq \label{eq. spin odd}
\Sigma^{\left(j\right)}_-=\frac{g^2}{1-\Sigma^{\left(j-2\right)}_-} ,\ \ \ \
\Sigma^{\left(j\right)}_+= \frac{g^2}{1-\Sigma^{\left(j-2\right)}_+}, \eeq
\beq \label{eq. Abox odd} A^{\left(j\right)}_{box}= g^2 A^{\left(j-1\right)}_{box}, \eeq
\beq \label{eq A odd} A^{\left(j\right)}=
\frac{A^{\left(j\right)}_{box}}{\left(1-\Sigma^{\left(j\right)}_+-\Sigma^{\left(j\right)}_-\right)^2}, \eeq
where $j=1, 3, 5 \ldots$.

\subsubsection{Even effective spins}

To find a generalized formula for even rows, it is useful to study
first the case of four rows.

The easiest way to calculate  the corresponding effective spin in
four rows  is by comparing it with the  two rows. We take the two
components of $\Sigma^{\left(3\right)}$ as the new fundamental
spins so that $\Sigma^{\left(4\right)}$ has the same form as
$\Sigma^{\left(2\right)}$ (\ref{eq. spin2 diagram}), but with the
$\Sigma_{\pm}^{\left(3\right)}$ replacing the $\sigma_{\pm}$: \beq
\label{eq. spin4} \Sigma^{\left(4\right)}=
\frac{g^2}{\left(1-\Sigma^{\left(3\right)}_+\right)\left(1-\Sigma^{\left(3\right)}_-\right)}.
\eeq

We can now calculate $A^{\left(4\right)}_{box}$. Diagrammatically,
it is
\[A^{\left(4\right)}_{box}=\barr{c}\mbox{\epsfig{figure=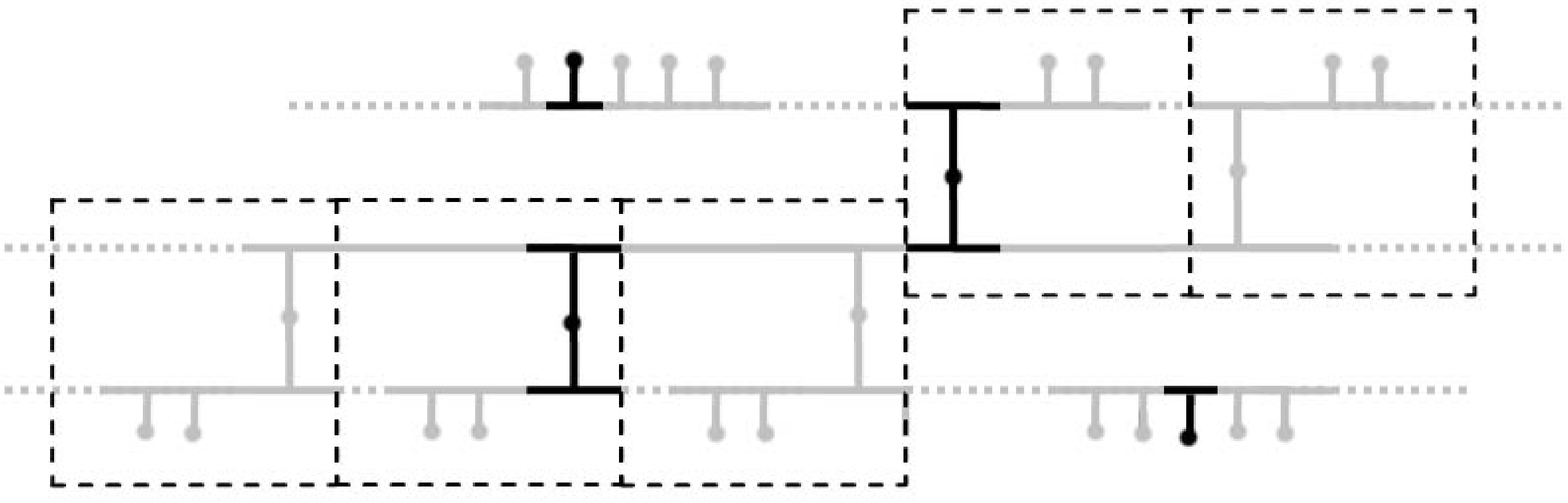,height=3cm}}\earr\]

Each box represents a $\Sigma^{(3)}$ spin and the number of boxes
ranges from zero to infinite. In the $A_{box}^{(4)}$ diagram, we
should guarantee the existence of at least two ``{\bf I}'' shapes
for the two intermediate slices. One of them is located inside one
of the bottom row of boxes and the other one inside one of the top
ones. In the above diagram they were indicated in black color. In
addition to these two ``{\bf I}'' shapes, we must also guarantee
the existence of at least one $\Sigma^{(1)}_-$ (i.e. $\sigma_-$)
in the initial time-slice and one $\Sigma^{(1)}_+$ (i.e.
$\sigma_+$) in the final time-slice, in order to connect initial
and final rings with the least number of connections between
slices. How can we choose the $\Sigma^{(1)}_-$ and
$\Sigma^{(1)}_+$ spins? A suggestion is that they are chosen among
the other gray $\Sigma^{(1)}$ spins inside the boxes. Although,
the suggestion is problematic because, in general, the two
$\Sigma^{(1)}$ spins may appear inside two boxes that are
different than the previously guaranteed ones. Therefore the
existence of such boxes, which support the two $\Sigma^{(1)}$
spins, should be guaranteed first. The existence of the box
requires the existence of its ``{\bf I}'' shape part. Turning on
more gray ``{\bf I}'' shapes, does not meet the initial motivation
of defining the notion of $A_{box}$, which was any necessary
connection between the initial and final rings such that at every
ring the existence of only {\it one spin} is guaranteed (due to
non-degeneracy constraint).

Another suggestion for supporting the existence of these two
$\Sigma^{(1)}$ spins is that some $\Sigma^{(1)}$ spins live
independently (with respect to boxes) on the initial and final
rings and among them the existence of one up and one down ones is
guaranteed. The idea is acceptable since it meets the condition of
the least number of guaranteed spins among the most arbitrary
configuration of connections between the initial and final rings
(which is 6 in this case).

Therefore the weight of $A^{\left(4\right)}_{box}$ can be
generally written:

\begin{eqnarray}
A^{\left(4\right)}_{box} &=& g^6xy \left[ \ \sum_{M=1}^{\infty}M
\left(\sigma_+\right)^M\ \cdot \ \sum_{N=1}^{\infty}N
\left(\sigma_-\right)^N\ \cdot \
\sum_{L=1}^{\infty}L\left(\Sigma^{\left(3\right)}_+\right)^L\
\cdot \
\sum_{K=1}^{\infty}K\left(\Sigma^{\left(3\right)}_-\right)^K \right]\nonumber \\
&=&
\frac{g^6xy}{\left(1-\sigma_{+}\right)^2\left(1-\sigma_{-}\right)^2\left(1-\Sigma^{\left(3\right)}_+\right)^2\left(1-\Sigma^{\left(3\right)}_-\right)^2}.
\label{eq.A4box}
 \end{eqnarray}

The final amplitude then is: \beq \label{eq. A4}
A^{\left(4\right)}= A^{\left(4\right)}_{box} \sum_{N=0}^{\infty}
\left(N+1\right) \left(\Sigma^{\left(4\right)}\right)^N =
\frac{A^{\left(4\right)}_{box}}{\left(1-\Sigma^{\left(4\right)}\right)^2}.
\eeq

Similarly, for the general even-$j$ rows, the effective spin, $A^{\left(j\right)}_{box}$ and general
amplitude $A^{\left(j\right)}$ are:
\beq \label{eq. spin_even} \Sigma^{\left(j\right)} =
\frac{g^2}{\left(1-\Sigma^{\left(j-1\right)}_+\right)\left(1-\Sigma^{\left(j-1\right)}_-\right)},\eeq

\beq
A^{\left(j\right)}_{box}
=\frac{g^{2\left(j-1\right)}xy}{\prod_{i=1}^{j-1}
\left(1-\Sigma^{\left(i\right)}_+\right)^2\left(1-\Sigma^{\left(i\right)}_-\right)^2 }
\label{Ajbox},
\eeq

\beq
A^{\left(j\right)}=\frac{A^{\left(j\right)}_{box}}{\left(1-\Sigma^{\left(j\right)}\right)^2},\label{Aj}
\eeq
where $j=2,4, 6 \ldots$.

\section{Renormalization Group}

Our aim in this section is to study the continuum limit of the statistical model we
defined. This is the limit in which the number of rows goes to infinity.
We do this by deriving a coarse-grained correlation function and searching
for critical behavior in which the Green's functions scale.
This is straightforward because the method we have used to solve the model already involves
defining and summing effective spin degrees of freedom.

It is important to note that the definition of effective spins
which then live on a one-dimensional chain, key to our solution,
works for 1+1 CDT because the spacelike and timelike couplings
scale differently and there is no coupling between timelike and
spacelike edges.

We denote the effective spin in the continuum limit by
$\Sigma^{\left(\infty\right)}$. In this limit, there is only one
component for the effective spin. Also, there is no difference
between even and odd slices.\footnote{Using the equation (\ref{eq.
spin_even}) for the only component of the effective spin, we see
the spin is invariant under the transformation $\Sigma^{(\infty)}
= \frac{g^2}{1-\Sigma^{(\infty)}}$. Therefore in the continuum
limit, there is no difference between odd and even effective
spins.} The effective spin weighs:

\beq \label{eq. spin infty} \Sigma^{\left(\infty\right)} =
\frac{g^2}{1-\frac{g^2}{1-\frac{g^2}{1-\frac{g^2}{1-\ddots}}}}
=\frac{1-\sqrt{1-4g^2}}{2}. \eeq
We now can apply a  renormalization group
transformation, acting on
a chain of infinite-row effective spins.
In the continuum limit, it is not necessary to consider
 the $A^{\left(\infty\right)}_{box}$ since it
is only one of the effective spins and the system is big enough to
consist of many effective spins. With this simplification, the
amplitude of a chain with $N$ spins is \beq \label{eq. RG
A^infty_N} A^{\left(\infty\right)}_{N} \simeq
\Sigma^{\left(\infty\right)}_{1} \Sigma^{\left(\infty\right)}_{2}
\Sigma^{\left(\infty\right)}_{3} \cdots
\Sigma^{\left(\infty\right)}_{N}. \eeq Rescaling of the Green's
function of the amplitude of gravity happens when the critical
exponent of dimensional rescaling factor is one \cite{ambj98}. We
start coarse-graining the above amplitude to \beq \label{eq. RG
A^infty_N/2} A^{\left(\infty\right)}_{\frac{N}{2}} \simeq
\Sigma^{\left(\infty\right)}_{1} \Sigma^{\left(\infty\right)}_{3}
\Sigma^{\left(\infty\right)}_{5} \cdots
\Sigma^{\left(\infty\right)}_{\frac{N}{2}}. \eeq The
renormalization group provides us with a specific parameter value
with which the amplitude is conformally invariant. This occurs
when three of finer spins are conformally equivalent to the
resealed amplitude of two coarser spins: \beq \label{eq. RG sigma
coarser finer}
\Sigma^{'\left(\infty\right)}_{1}\Sigma^{'\left(\infty\right)}_{3}
= 2\ \Sigma^{\left(\infty\right)}_{1}
\Sigma^{\left(\infty\right)}_{2}\Sigma^{\left(\infty\right)}_{3} ,
\eeq where $2$ is the dimensional rescaling factor.

Substituting equation (\ref{eq. spin infty}) in (\ref{eq. RG sigma
coarser finer}) we obtain
\[ \left(\frac{1-\sqrt{1-4g'^{2}}}{2}\right)^{2} = 2\
\left(\frac{1-\sqrt{1-4g^{2}}}{2}\right)^{3}.
\]

We now see that $g'$ and $g$ have a non-linear relation:

\beq \label{eq. RG3} g'=\pm \frac{1}{2} \sqrt{1-\left(1 -
\left(1-\sqrt{1-4g^2}\right)^{\frac{3}{2}}\right)^2} =: \pm
f\left(g\right), \eeq

or, graphically, \footnote{There is also another possible relation
$g'=\pm \frac{1}{2} \sqrt{1-\left(1 +
\left(1-\sqrt{1-4g^2}\right)^{\frac{3}{2}}\right)^2}$, whose fixed
point is at $g=0$.}

\begin{center}
\includegraphics[scale=.45]{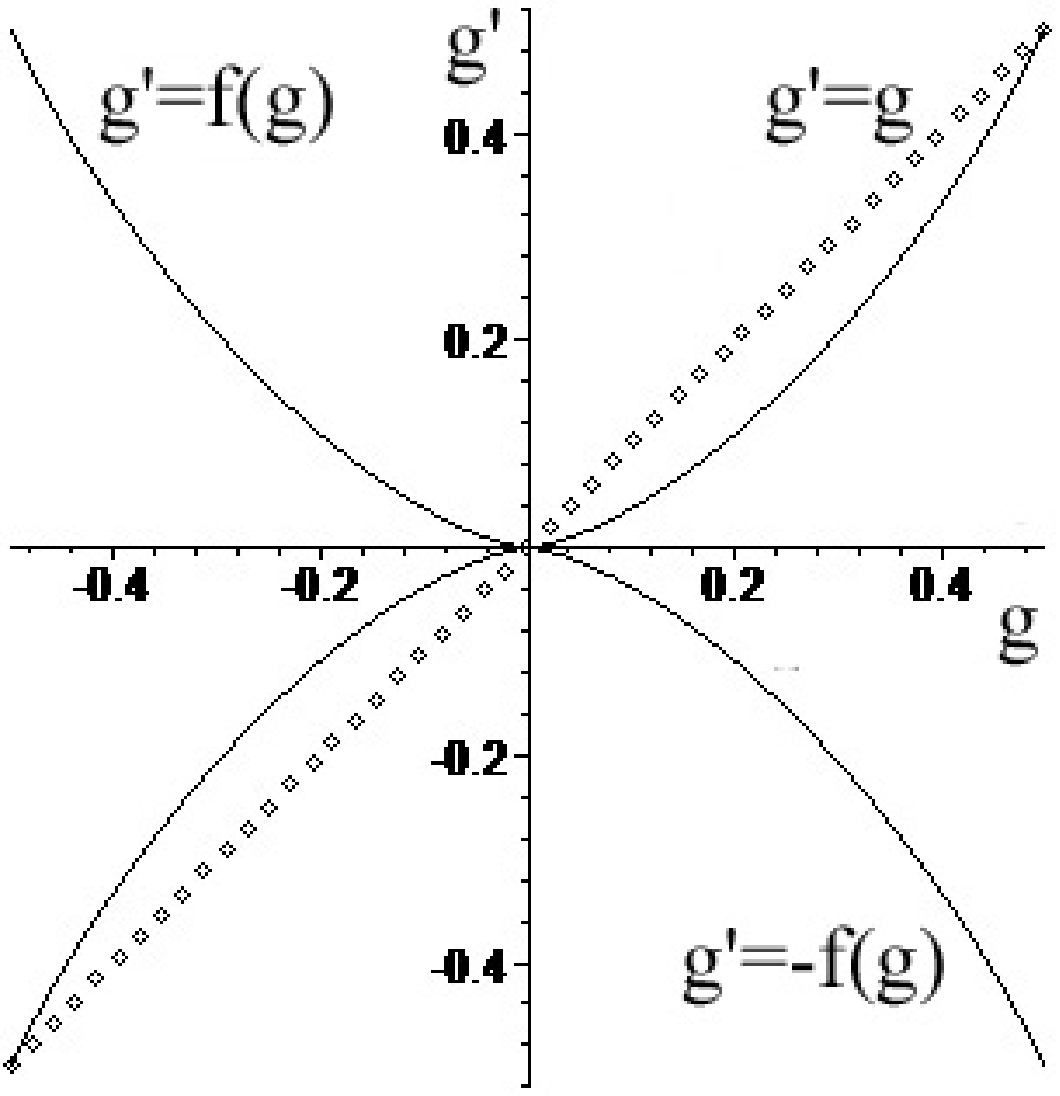}
\end{center}

We can iterate this coarse-graining to produce a coarser
one-dimensional chain. One can see that, after a few iterations,
the triangular coefficients approach the fixed points $g=0$ and
$g=\pm \frac{1}{2}$. This result agrees with the continuum limit
of the Causal Dynamical Triangulation model\cite{ambj98}.

Inserting $\Sigma^{\left(\infty\right)}$ in equations (\ref{Ajbox}) and (\ref{Aj}), we find that
\begin{eqnarray}
 A^{\left(\infty\right)}_{box}&&=\nonumber\\
&& \lim_{j\rightarrow\infty}
\frac{g^{2\left(j-1\right)}xy}{[\left(1-\sigma^{-}\right)\left(1-\sigma^{+}\right)\left(1-\Sigma^{\left(3\right)-}\right)\left(1-\Sigma^{\left(3\right)+}\right)\ldots
\left(1-\Sigma^{\left(j-1\right)-}\right)\left(1-\Sigma^{\left(j-1\right)+}\right)]^2} \nonumber\\
&&
 \label{eq. abox infty}
 \end{eqnarray}
and, therefore, the amplitude in the continuum limit
is:
\begin{eqnarray}
 A^{\left(\infty\right)}&=&\frac{A^{\left(\infty\right)}_{box}}{\left(1-\Sigma^{\left(\infty\right)}\right)^2}\nonumber\\
& =&
\frac{4A^{\left(\infty\right)}_{box}}{\left(1+\sqrt{1-4g^2}\right)^2}.
\label{eq. A infty}
\end{eqnarray}
The denominator in the region of convergence (where $|g|\leq 0.5$,
$|x| \leq 1$ and $|y|\leq 1$) is always positive and non-zero.

Note that the denominator of equation (\ref{eq. abox infty}) is an
infinite product of terms, each term  defined recursively from the
previous terms.    Since the  limit of the effective spins is
$\Sigma^{\left(\infty\right)}$,  we can approximate
$A^{(\infty)}_{box}$ with its limit: \beq
A^{\left(\infty\right)}_{box} \leq \lim_{j\rightarrow \infty}
\frac{g^{2\left(j-1\right)}xy}{\left(1-\Sigma^{\left(\infty\right)}\right)^{4\left(j-1\right)}}.
\label{eq._Abox_infty approx} \eeq

\section{Discussion}

In this paper we constructed a spin system with constraints
that provides a dual description of the 1+1 Causal Dynamical Triangulation model
of \cite{ambj98}.  By inventing a notion of effective spins, we were able to
solve the model.
We should emphasize that the fact the model is solvable
is not new, and the solution we find is completely equivalent to that of
the original paper of Ambjorn and Loll \cite{ambj98}. However, the dual spin model gives an alternative way of understanding what it means to define a sum or path integral
over causal structures. As such, we expect it may be useful in higher dimensions
and when matter is included.

In closing we make a few observations that may be useful for future work in this direction.

\begin{itemize}

\item{} Our effective spins are quite different from the ones used in coarse-graining
of standard spin systems.  Here the effective spins really carry out the sum over spin lattices necessary in a quantum gravity model.

\item{}There is an analogy between curvature and the hamiltonian of an
antiferromagnetic spin system.  The reason is that a triangulated
2d flat spacetime corresponds to the case in which the triangles
in each time slice alternate between up and down triangles. In the
dual spin system this is a configuration in which up spins
alternate with down spins. Also, the causality conditions require
that the spins alternate between up and down in the timelike
direction.  Hence, if the ground state of the gravity theory for
small cosmological constant is locally approximated by flat
spacetime, this will correspond to a configuration of spins which
alternates in both the space and time directions.  This suggests
that the ground state of the gravity system should resemble the
ground state of a two-dimensional antiferromagnetic system.
Whether this generalizes in higher dimensions will be investigated
in future work.

\item{}In the CDT model, a Euclidean continuation is made which maps the quantum
gravity system in $1+1$ dimensions to a classical statistical system in $2$ dimensions.
We note that it may be possible instead to use the method described here to map
the quantum gravity system to a one-dimensional quantum spin chain and solve
it directly as a quantum mechanical system.

\end{itemize}

\section*{Acknowledgments}
We are  grateful to Lee Smolin for comments and suggestions on the
manuscript. We also thank Ali Tabei for very useful discussions
clarifying the statistical behavior of spin systems, Karol
\.{Z}yczkowski for helpful conversations and Hal Finkel for
comments on the manuscript.

\end{document}